\documentstyle[prl,preprint,aps,psfig]{revtex}
\begin{document}
\draft
\title{Fermi-sea-like correlations in a partially filled Landau level}
\author{R. K. Kamilla$^{1,2}$, J. K. Jain, and S. M. Girvin$^3$}
\address{$^1$Department of Physics, State University of New York
at Stony Brook, Stony Brook, New York 11794-3800}
\address{$^2$
Department of Physics, MIT, Cambridge, Massachussetts 02139}
\address{$^3$
Department of Physics, Indiana University, Bloomington, Indiana 47405}
\date{\today}
\maketitle
\begin{abstract}

The pair distribution function and the static structure factor
are computed for composite fermions. Clear and robust 
evidence for a $2k_F$ structure is seen in a range of filling 
factors in the vicinity of the half-filled Landau level. 
Surprisingly, it is found that filled Landau levels of composite
fermions, i.e. incompressible FQHE states, bear a stronger 
resemblance to a Fermi sea than do filled Landau levels of electrons. 

\end{abstract}

\pacs{71.10.Pm,73.40.Hm}

\section{Introduction}

A central feature of the composite fermion (CF) theory 
\cite{book,Jain,HLR} of the quantum Hall effect 
is that the interaction energy of electrons is 
transmuted into the kinetic energy of composite fermions. In 
other words, while the original problem had no kinetic energy 
(more precisely, a constant kinetic energy) but only the 
interaction energy, the kinetic energy is the important 
energy of composite fermions, the interaction between 
which is weak and can be neglected altogether to a good first 
approximation. The kinetic energy of composite fermions 
manifests itself in the form of Landau levels (LLs) of 
composite fermions, resulting in the fractional quantum Hall
effect (FQHE), and, in an extreme limit, as the Fermi sea 
of composite fermions at the half filled Landau level.

In this work, we will compute the pair distribution 
function and the static structure factor
for filled Landau levels of composite fermions, corresponding
to FQHE states at $\nu=n/(2n+1)$, as well as
for composite fermions at zero effective magnetic field,
using the recently developed lowest-LL representation of the CF 
wave functions that enables a treatment of large CF systems 
\cite {JK}. The qualitative behavior of the results
provides a striking evidence for  
Fermi-sea-like correlations in a broad range of filling
factors near the half-filled Landau level.

The results obtained below are 
valid to the extent the CF wave functions are. The 
CF wave functions have been compared extensively to the exact 
eigenstates, available numerically for systems containing 
$\sim 10-12$ particles, and found to have close to 100\% overlap with
them \cite {Dev,Rezayi,JK,Haldanehalf}. 
It is quite plausible that the excellent agreement 
will persist to larger systems and also for other incompressible
FQHE states, suggesting that the CF wave functions should be
reliable up to 9/19, which is the fraction closest to 1/2 where
experimental indications for FQHE 
have been observed \cite {Du}. There is sound experimental 
evidence that the CF description is valid for the compressible state
at $\nu=1/2$ as well \cite {book,semi}. Here, the CF wave 
functions are expected to capture the short-distance behavior of
the composite-Fermi sea state but may not to be very useful for 
subtleties associated with the low-energy long-distance physics;
unlike the FQHE states, the composite
Fermi-sea is likely to be susceptible to perturbations 
due to the absence of a gap. 
Fortunately, however, we will see that it is not necessary to go 
all the way to 1/2 to see Fermi-sea-like correlations; they 
are manifest even in incompressible FQHE states away from 1/2,
indicating that a composite-Fermi-sea description of the half-filled
LL state is valid, at least as a first order description. 
The issues concerning subtle and 
asymptotic deviations from a {\em perfect} Fermi liquid, 
which can surely be expected given that there was no
reason to expect any kind of Fermi sea in the first place, are  
beyond the scope of the present study.

Another widely used approach for an investigation of the CF state
is based on a Chern-Simons field theory  
\cite {HLR,Lopez,Shankar}, useful for a  
description of the low-energy long-distance physics 
of the compressible composite-Fermi sea at $\nu=1/2$, which has 
been found to be reasonably stable against perturbations 
\cite {HLR}. The complementary approach of the 
present work provides support to this conclusion.

\section{Non-interacting electrons}

We will be concerned with two (related) quantities, the 
pair distribution function, $g(r)$, and the static structure factor,
$S(k)$, of composite fermions. 
Before proceeding to the study of composite fermions, we 
discuss the behavior for non-interacting electrons, which will
be useful when we come to composite fermions.

The pair distribution function $g(r)$ is equal to the probability
of finding two particles at a distance $r$, normalized so that it 
approaches unity for large $r$. Two expressions for this
quantity are 
\begin{equation}
g(r)=\frac{1}{\rho N} <\sum_{i\neq j} \delta(r-r_i+r_j)>\;,
\label{gr}
\end{equation}
and
\begin{equation}
g(|r_1-r_2|)=\frac{N(N-1)}{\rho^2} \int ... \int d^2r_3 ... d^2r_N 
|\Psi(r_1,r_2,...,r_N)|^2\;,
\end{equation}
where the wave function $\Psi(r_1,r_2,...,r_N)$ is assumed to be
normalized to unity.

For single-Slater-determinant (Hartree-Fock) states, $g(r)$ can be 
evaluated straightforwardly by writing it as
\begin{equation}
g(|r_1-r_2|)=\frac{1}{\rho^2}\sum_{j,k}
|\eta_j(r_1) \eta_k(r_2)-\eta_k(r_1)\eta_j(r_2)|^2\;,
\end{equation}
where $\eta$ are single electron states  
and the sum is over all occupied states. 
For the free Fermi sea at zero magnetic field,  this gives 
\begin{equation}
g(r)=1-\left[ \frac{2}{k_F r} J_1(k_F r)\right]^2\;,
\end{equation}
which, for large $rk_F$, behaves as
\begin{equation}
g(r)=1-\pi^{-1} \left(\frac{2}{rk_F}\right)^3 \sin^2(\frac{\pi}{4}
-rk_F)\;.
\end{equation}
For the state with $n$ filled Landau levels, one obtains  
\begin{equation}
g(r)=1-\frac{1}{n^2}e^{-\frac{(rk_F)^2}{4n}}
\left[ L^1_{n-1}(\frac{(rk_F)^2}{2n})\right]^2\;.
\end{equation}
The Fermi sea result can be obtained from this 
by taking the limit $n\rightarrow \infty$ and using
$$ \lim_{n\rightarrow \infty}\;\frac{1}{n} 
L_n^1(\frac{x}{n})=x^{-1/2} J_1(2\sqrt{x})\;.$$
Fig. \ref{fig:FS_fig4} shows how the $g(r)$ of filled LL states 
evolves into that of the Fermi sea.

The static structure factor $S(k)$ is given by 
\begin{equation}
S(k)=\frac{1}{N} \Pi(k)-N \delta_{k,0}\;,
\label{sk}
\end{equation}
where  
\begin{equation}
\Pi(k)\equiv <\rho(k)\rho(-k)>\;
\end{equation}
\begin{equation}
\rho(k)=\sum_j e^{i{\bf k}.{\bf r}_j}\;,
\end{equation}
and $<>$ denotes ground state expectation value.
It can also be obtained from the dynamical structure factor 
$S(k,\omega)$ as
\begin{equation}
S(k)=\int_0^{\infty} \frac{d\omega}{\pi} S(k,\omega)\;.
\end{equation}
$S(k,\omega)$, in turn, is related to the imaginary part of
the inverse dielectric function, which, for non-interacting
electrons, leads to 
\begin{equation}
S(k)=-\rho^{-1} \int_0^{\infty} \frac{d\omega}{\pi} \Im[\Pi^0(k,\omega)]\;,
\end{equation}
where $\Pi^0(k,\omega)$ is the density-density 
response  function. For two-dimensional spinless electrons in 
a Fermi sea \cite {Stern},
\begin{equation}
\Pi^0(k,\omega)= \frac{\rho}{2\epsilon_F}\frac{2k_F}{k}
[2 \frac{k}{2k_F} -(a_+^2-1)^{1/2}+(a_--1)^{1/2}]\;,
\end{equation}
where 
\begin{equation}
a_{\pm}=\frac{\omega+i\epsilon}{kv_F} \pm  \frac{k}{2k_F}\;,
\end{equation}
with the square root is chosen to be on the branch with positive
imaginary part. The static structure factor can be evaluated and found
to be
\begin{equation}
S(k)  =  \frac{2}{\pi}\left[\frac{k}{2k_F} \sqrt{1-(\frac{k}{2k_F})^2}
\right] +
\arcsin(\frac{k}{2k_F})\;, \;\;\; (k<2k_F)\;
\end{equation}
\begin{equation}
=1\;,\;\;\; (k \geq 2k_F).
\end{equation}
At small $k$,
\begin{equation}
S(k)\approx \frac{4}{\pi}[\frac{k}{2k_F} -\frac{1}{6} (\frac{k}{2k_F})^3]\;.
\end{equation}
Note that there is no discontinuity at $k=2k_F$ in either $S(k)$ or 
its first derivative (although there is one in the second derivative).

The correlation functions $g(r)$ and $S(k)$ are related by 
Fourier transform as
\begin{equation}
S(k)-1=\rho \int d^2 r \; e^{i{\bf k}.{\bf r}} (g(r)-1)\;.
\end{equation} 
It is also important to note that both $g(r)$ and $S(k)$ 
are properties of the ground state.

\section{Composite fermion wave functions}

We will employ the spherical geometry 
in which $N$  electrons move on the two-dimensional
surface of a sphere \cite{Haldane,WuYang} under the influence
of a radial magnetic field $B$ originating from a magnetic 
monopole of strength $Q$ at the center, which corresponds to a
total flux of $2Q\phi_0$, where $\phi_0=hc/e$ is the flux quantum. 
An ideal limit of zero transverse width, no disorder, 
vanishing LL mixing, and fully-polarized electrons will be assumed,
allowing us to consider spinless electrons confined to the lowest 
LL moving on  a disorderfree two-dimensional surface. 

The CF theory maps the interacting electrons at $Q$ 
into weakly interacting fermions at a reduced magnetic field, 
corresponding to a monopole strength
\begin{equation}
q=Q-p(N-1)\;,
\end{equation}
where $p$ is an integer.
The wave function of composite fermions at $q$ (often denoted by
$q^*$ or $q^{CF}$; the superscript will be omitted here for
convenience) is related to the wave function of {\em 
non-interacting} electrons at $q$,
denoted by $\Phi$ (which is in general a linear superposition of 
Slater determinants) as
\begin{equation}
\Phi^{CF}={\cal P}\Phi_1^{2p}\Phi\;,
\end{equation}
where $\Phi_1$ is the wave function of the lowest filled Landau level
and ${\cal P}$ is the lowest LL projection operator. 

In general it is quite difficult to explicitly evaluate
the projection operator ${\cal P}$ acting on the full 
wave function for large numbers of particles. In Ref. \cite{JK}
a new and much more convenient approach to this problem was 
developed. Here, first the Jastrow factor is written as
\begin{equation}
\Phi_1^{2p} = \prod_j J^p_j,
\end{equation}
where
\begin{equation}
J_j \equiv \prod_k(u_jv_k - v_ju_k)\exp\left[\frac{i}{2}(\phi_j +
\phi_k)\right]
\end{equation}
and $\phi_j$ is the $j$th azimuthal angle.  Thus $\Phi_1^{2p}$ can
be replaced by a factor of $J^p_j$ in every element of the $j$th column of
the determinantal matrix. Then, the projection on lowest LL is 
accomplished by separately projecting each element. This 
modified wave function is not identical to that used previously, but
very similar and has similar correlations built in. It should be
noted that there is nothing to choose from, {\em a priori}, between
different ways of obtaining a lowest LL wave function from the
unprojected CF wave function.
It is necessary in each case, however, to test the
validity of the lowest LL wave function against exact solutions
known from numerical diagonalization studies on finite systems.
Ref. \cite {JK} carried out such comparisons for systems with up to
12 particles, showing that the new wave functions are also 
extremely accurate. 

In this approach \cite {JK}, $\Phi^{CF}$  
can be obtained from $\Phi$ by replacing the single-electron eigenstates, 
$Y_{q,n,m}(\Omega)$, by single-CF states, $Y^{CF}_{q,n,m}(\Omega)$.
Here $n=0,1,2,...$ is the LL index, $l=q+n$ is the orbital
angular momentum, $m=-q-n, -q-n+1, ..., q+n$
labels the degenerate states in the $n$th LL,
and $\Omega$ represents the angular coordinates
$\theta$ and $\phi$ of the fermion.
$Y^{CF}_{q,n,m}$ is given by
\begin{eqnarray}
&&Y_{q,n,m}^{CF}(\Omega_j) =
N_{qnm} (-1)^{q+n-m} \frac{(2Q+1)!}{(2Q+n+1)!}
u_j^{q+m}   v_j^{q-m}  \nonumber \\
&&
\sum_{s=0}^{n}(-1)^s {{n \choose s}} {{ 2q+n \choose q+n-m-s}}
\; u_j^s \;    v_j^{n-s}  \; \left[ \left({\frac{\partial}{\partial u_j}}
\right)^s 
\; \left({\frac{\partial}{\partial v_j}}\right)^{n-s} J_j^p\right]\;
\end{eqnarray}
where
\begin{equation}
J_j=\prod_{k}^{'} (u_j v_k-v_j u_k)\;,
\end{equation}
\begin{equation}
u\equiv \cos(\theta/2)\exp(-i\phi/2)\;,
\end{equation}
\begin{equation}
v\equiv \sin(\theta/2)\exp(i\phi/2)\;,
\end{equation}
the prime denotes the condition $k\neq j$, and $N_{qnm}$ is 
the normalization constant (not relevant for the calculation
below, which involve only single-Slater determinant states).
The binomial coefficient ${{\alpha \choose \beta}}$ is to be set
to zero if either $\beta>\alpha$ or $\beta<0$.

In order to evaluate the derivatives, we find it convenient to 
write them as follows:
\begin{equation}
\left({\frac{\partial}{\partial u_j}}\right)^s
\; \left({\frac{\partial}{\partial v_j}}\right)^{n-s} 
J_j^p= J_j^p \left[ \overline{U}_j^s \overline{V}_j^{n-s} 1 \right]
\end{equation}
where 
\begin{equation}
\overline{U}_j=J_j^{-p} \frac{\partial}{\partial u_j} 
J_j^{p}=p \sum_{k}^{'}\frac{ v_k}{
u_j v_k - v_j u_k} + \frac{\partial}{\partial u_j}\;\;,
\end{equation}
\begin{equation}
\overline{V}_j= J_j^{-p} \frac{\partial}{\partial uv_j} 
J_j^{p}=p\sum_{k}^{'}\frac{
-  u_k}{u_j v_k - v_j u_k} + \frac{\partial}{\partial v_j}\;.
\end{equation}
For a given $n$, the 
explicit analytical form of the derivatives
is used in the evaluation of the wave function.

In short, $Y^{CF}_{q,n,m}$ can be interpreted as the
single-CF wave function, from which many-CF wave functions are 
constructed in complete analogy to how many-electron wave functions
are constructed from single electron wave functions.
It should be noted that the wave function of a  
composite fermion actually
depends on the coordinates of all other particles, as a result of 
the strongly correlated nature of the problem.

\section{Results for composite fermions}

We now compute $g(r)$ and  $S(k)$ for composite fermions.
The former is given by the probability of finding a pair of
composite fermions at a distance $r$, while the latter is 
related to the equal time density-density correlation function
for composite fermions. Since the positions and the density 
operator are the same for electrons and composite fermions,
$g(r)$ and $S(k)$ for composite fermions are the same as those 
of electrons, with the expectation values evaluated in the
CF wave functions.

Two strategies will be employed for approaching the 
composite Fermi sea state. In one, we will obtain  
thermodynamic estimates for $g(r)$ and  
$S(k)$ for FQHE states at $\nu=n/(2n+1)$ 
for $n\leq 6$, a result interesting in its own right,
and look for systematic behavior leading to the composite 
Fermi sea, obtained in the limit $n\rightarrow \infty$. 
The FQHE state at
\begin{equation}
\nu=\frac{n}{2pn+1}\;
\end{equation}
has $n$ filled LLs of composite fermions and occurs at
effective monopole strength
\begin{equation}
q=\frac{N-n^2}{2n}\;,
\label{q}
\end{equation}
or at real monopole strength
\begin{equation}
Q=(p+\frac{1}{2n})N-(p+\frac{n}{2})\;.
\end{equation}
In the thermodynamic limit $N\rightarrow \infty$ the filling
factor $\nu\equiv N/(2Q)=n/(2pn+1)$.
In various studies, the CF wave functions have been tested for
2/5, 3/7, 2/3, and 3/5 for
systems of up to 12 particles, and the energies have been found
to be very accurate \cite {Dev,JK}.
We note that the CF state at $\nu=1/3$ is identical to the Laughlin
state \cite {Laughlin}, and that the non-interacting electron
wave function $\Phi$ becomes 
the Fermi sea state in the limit of $n\rightarrow\infty$.

The second approach will be to increase the number of 
particles while staying at a vanishing effective magnetic 
field of composite fermions. Zero effective flux ($q=0$) implies
\begin{equation}
Q=p(N-1)\;
\end{equation}
giving the compressible Fermi sea state
$\nu=1/(2p)$ in the limit $N\rightarrow \infty$.
The CF wave functions at this flux have been
studied by Jain and collaborators \cite {Dev},
by Rezayi and Read \cite {Rezayi}, and more recently by
Haldane in periodic geometry \cite {Haldanehalf}.
We will consider only the ``filled shell" systems with $N=n^2$, 
because the ground state here is a single Slater determinant,
making computations easier, and also because it is spatially 
uniform (with $L=0$) making $g({\bf r_1},{\bf r_2})$ a function 
of $|{\bf r_1}-{\bf r_2}|$ alone. Note that the 
zero-effective-flux system with $N=n^2$ is also
the smallest member of the incompressible $n/(2pn+1)$ FQHE state,
according to Eq.~(\ref{q}); however, the compressible 
$\nu=1/(2p)$ will be obtained in the limit of 
$N \rightarrow \infty$.

\subsection{Pair correlation function}

\begin{figure}
\centerline{\psfig{file=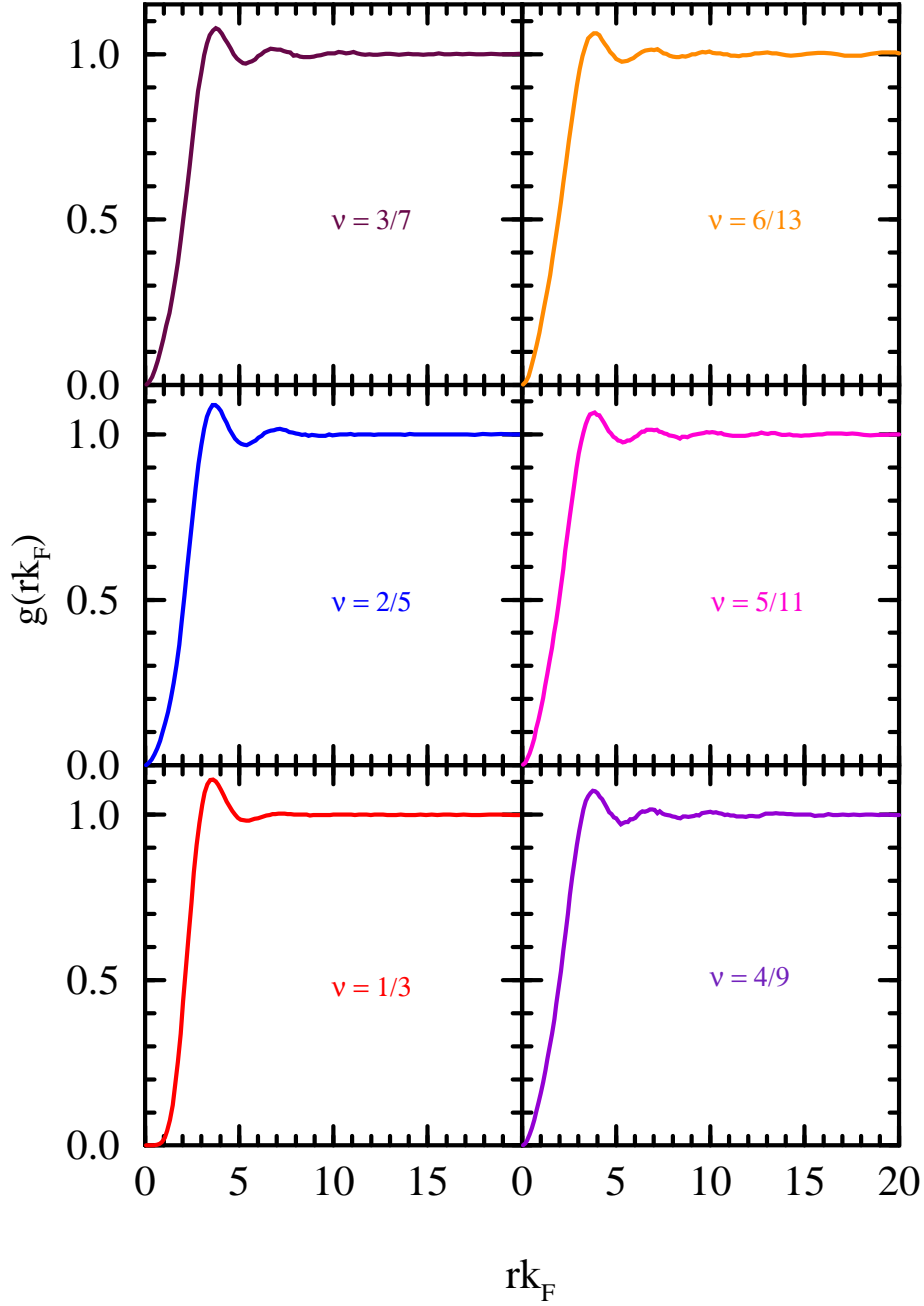,width=6in,angle=0}}
\caption{ Pair distribution functions for several FQHE states
at $\nu=n/(2n+1)$, corresponding to $n$ filled Landau levels 
of composite fermions. $N=$ 50, 54, 60, 60, 60, and 54 
particles have been used for $n=$ 1, 2, 3, 4, 5, and 6, 
respectively. The statistical error in Monte Carlo is of the
order of the noise on the curves. }
\label{fig:FS_fig2}
\end{figure}

\begin{figure}
\centerline{\psfig{file=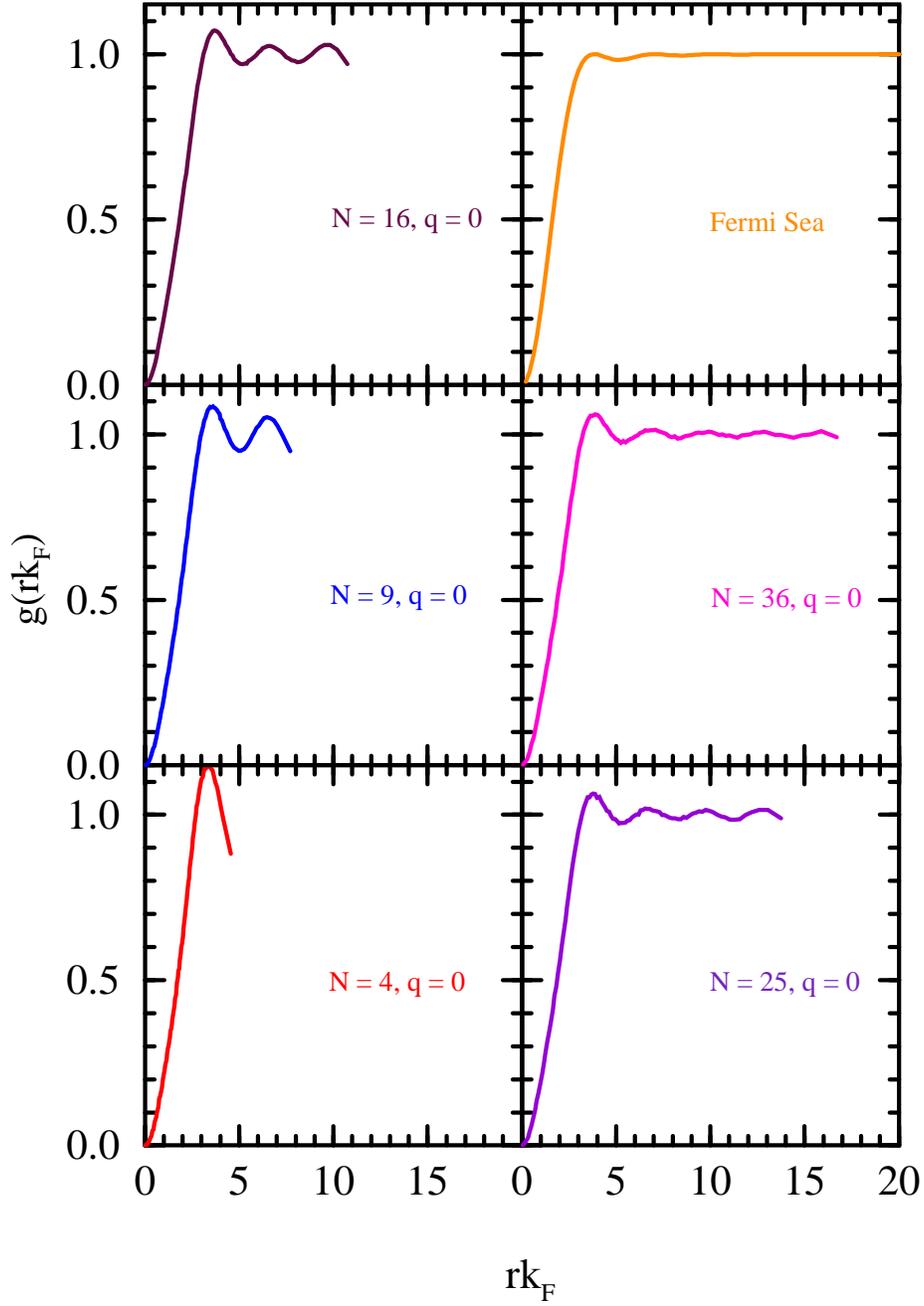,width=6in,angle=0}}
\caption{Pair distribution functions for several systems of 
composite fermions at zero effective magnetic field.
The number of particles is shown on the figures in each case. }
\label{fig:FS_fig3}
\end{figure}

Eq.~(\ref{gr}) is used for a Monte Carlo evaluation of 
the pair-correlation function of composite fermions,
with the distance $r$ defined to be the arch distance between the 
electrons. We have carried out a systematic study as a function 
of $N$ for FQHE states at $\nu=n/(2n+1)$ for $n\leq 6$. Fig. \ref{fig:FS_fig2} 
shows $g(r)$ for the largest system studied at each fractions
\cite {Morf}.  There is no appreciable $N$ dependence for 1/3 
to 4/19, and also for 5/11 and 6/13 except for the last 
oscillation. Fig. \ref{fig:FS_fig3} shows $g(r)$  for $q=0$ CF states for
up to 36 particles. The result for 9 particles is virtually 
indistinguishable from that obtained by exact diagonalization
\cite {Rezayi}, not surprising given that the CF 
wave function has a large overlap (0.99938 \cite {Dev,Rezayi}) 
with the exact ground state. 

\begin{figure}
\centerline{\psfig{file=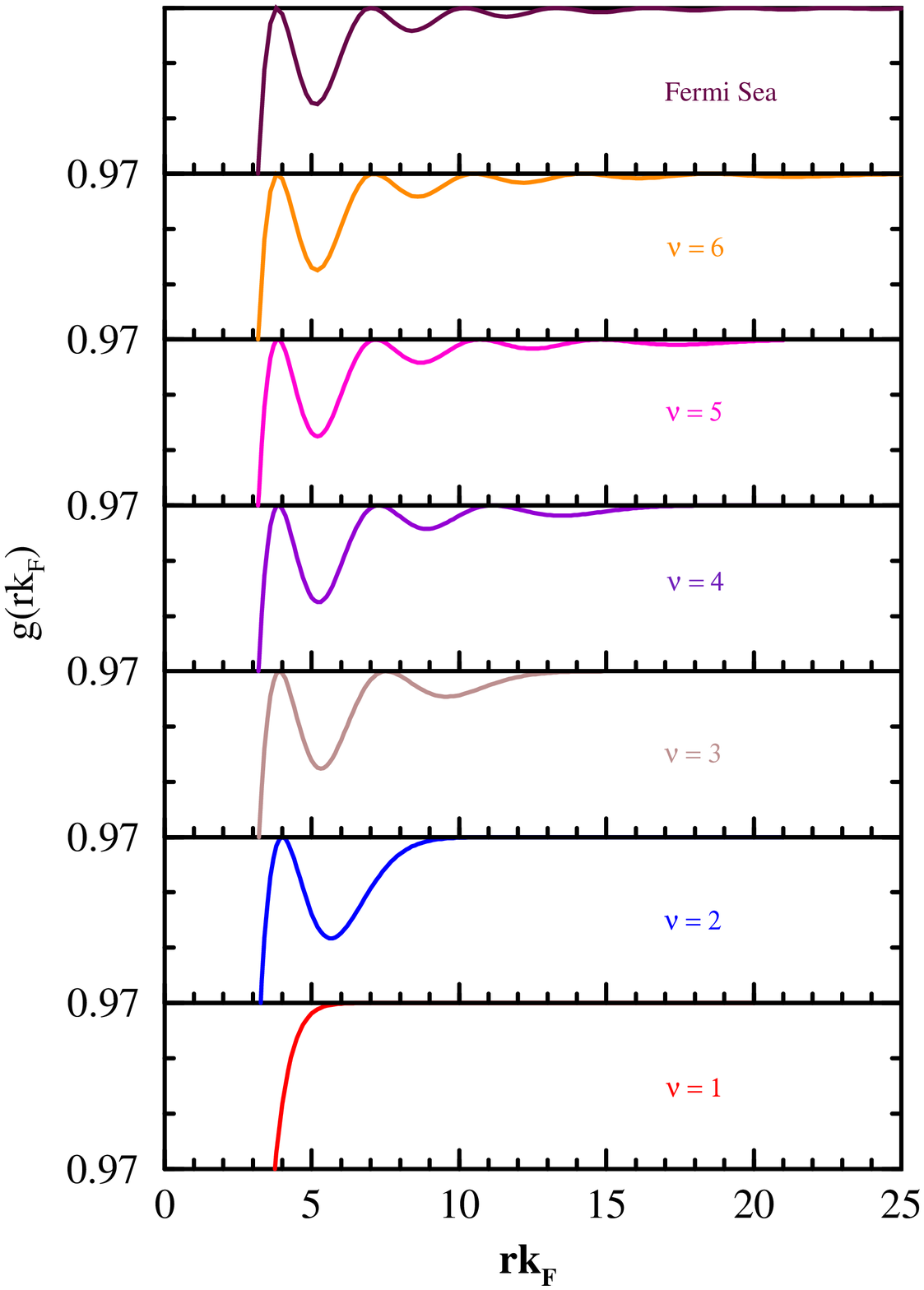,width=6in,angle=0}}
\caption{Pair distribution function for non-interacting electrons 
at integer filling factors ($\nu=n$) and also for the Fermi sea.}
\label{fig:FS_fig1}
\end{figure}

\begin{figure}
\centerline{\psfig{file=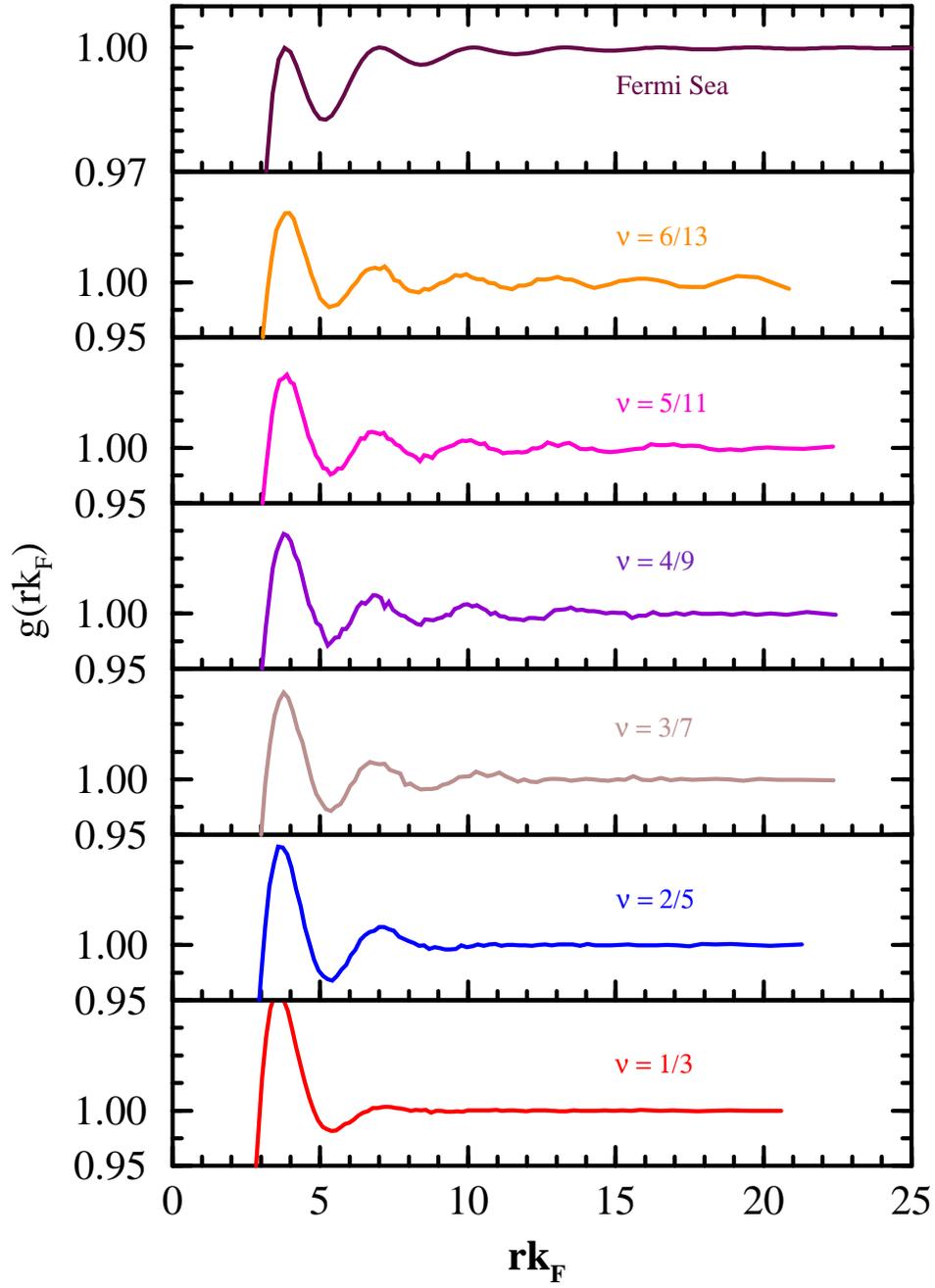,width=6in,angle=0}}
\caption{A close-up of oscillations in $g(r)$ for filled 
LLs of composite fermions. The top panel shows the $g(r)$
for non-interacting Fermi sea at zero magnetic field (same as 
in Fig. \protect \ref{fig:FS_fig2}).}
\label{fig:FS_fig4}
\end{figure}

\begin{figure}
\centerline{\psfig{file=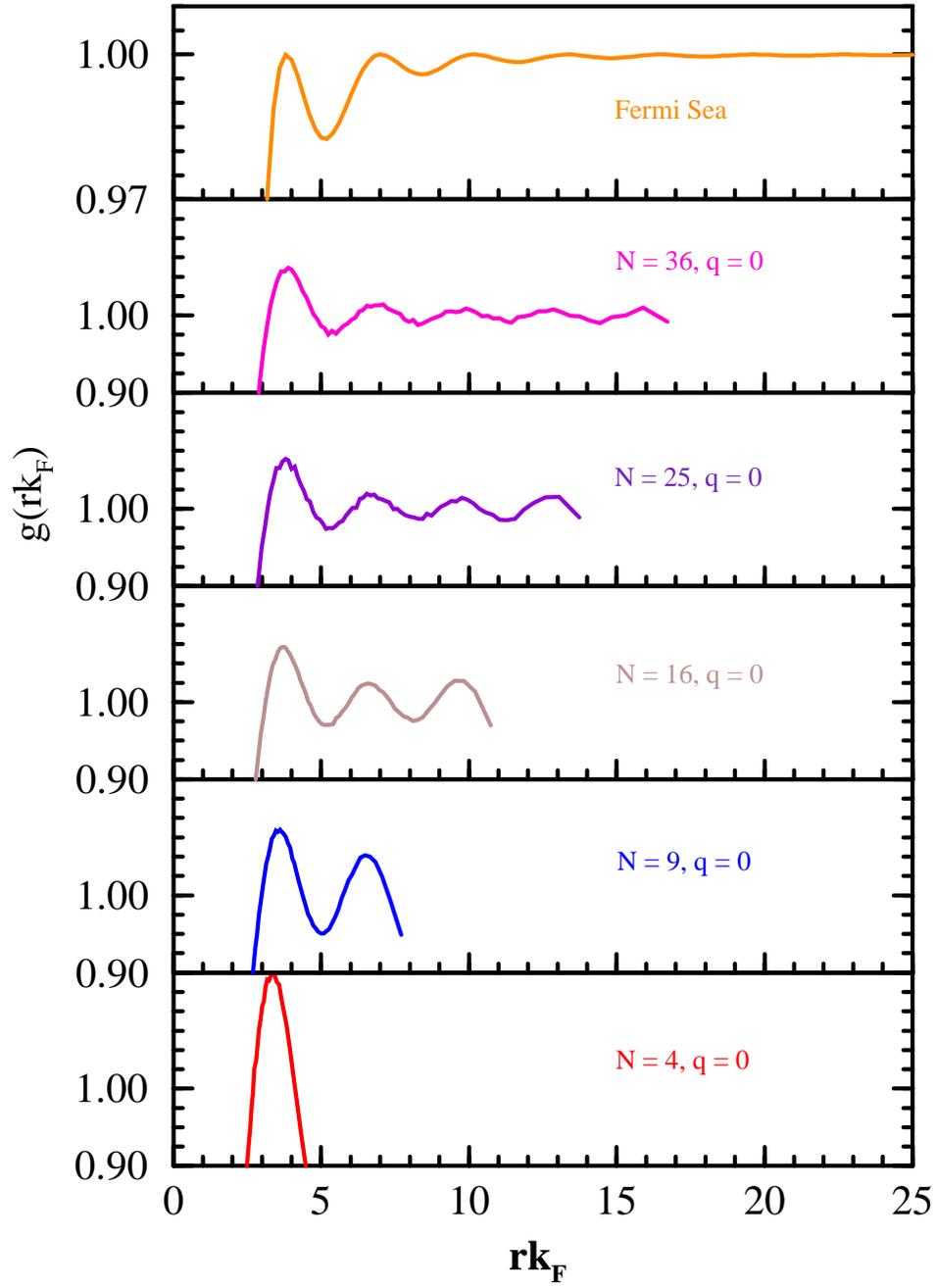,width=6in,angle=0}}
\caption{ Same as Fig. \protect \ref{fig:FS_fig4} but for composite fermions at 
vanishing effective magnetic field. }
\label{fig:FS_fig5}
\end{figure}

It is illuminating to contrast the oscillations
in $g(r)$ of the FQHE and the zero-effective-flux states with 
the $g(r)$ of a filled Fermi sea, as done in Figs. \ref{fig:FS_fig4} and \ref{fig:FS_fig5}.
An evolution into a Fermi-sea-like state is evident upon moving 
to larger values of $n$ along the $n/(2n+1)$ FQHE states, or upon 
increasing $N$ at zero effective flux, strongly indicating 
the presence of a Fermi-sea-like state at $\nu=1/2$.
In fact, the LLs of composite fermions bear 
a closer resemblance to a Fermi sea than do the LLs of electrons
in two respects. First, more oscillations occur; for example, 
two peaks can be seen right at 1/3, the state with
one filled LL of composite fermions, in contrast to the one filled
LL of electrons, the $g(r)$ of which does not have any oscillations.
Second, the peak positions are much better correlated with those 
of the Fermi sea in Figs. \ref{fig:FS_fig4} and \ref{fig:FS_fig5} than in Fig. \ref{fig:FS_fig1}.
This implies that the formation
of electron LLs is a stronger perturbation on the Fermi sea of
electrons than the formation of CF-LLs is on the Fermi sea of 
composite fermions, or in other words, that the 
Fermi sea of composite fermions is quite robust.

\begin{figure}
\centerline{\psfig{file=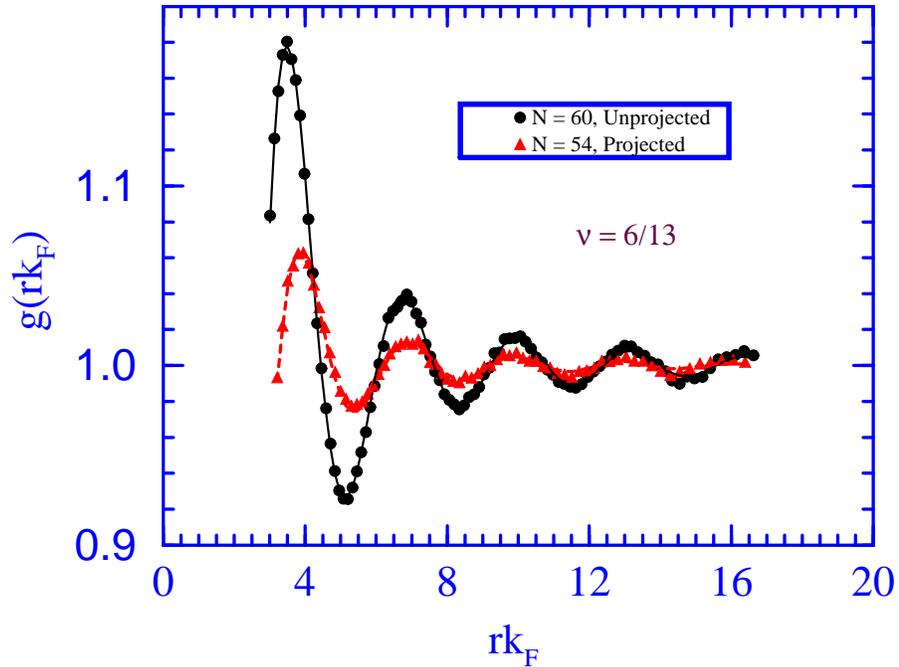,width=6in,angle=-90}}
\caption{ The pair correlation function, 
$g(r)$, for projected as well as the unprojected
CF state at $\nu=6/13$. $N=54$ (60) particles were used for the 
projected (unprojected) CF wave function. The points are the
calculated values and the curves are fits according to 
Eq.~(\protect \ref{fit}) with $\alpha= $ 2.75 (2.37), $\beta=$ 2.11 (1.98),
$\phi=$ 0.58 (-0.64), and $A=$ 2.71 (3.63) for the projected
(unprojected) CF state. }
\label{fig:FS_fig6}
\end{figure}

For a more quantitative analysis of the numerical
results we investigate the behavior of $g(r)$ at large $r$. 
For the FQHE states the oscillations are suppressed beyond 
a certain distance, as evident in Fig. \ref{fig:FS_fig4}  for 1/3, 2/5, 
3/7, and 4/9, as expected from the presence of a gap in these 
systems. We are interested in the power law fall off of 
$g(r)$ in the composite-Fermi-sea state.  
As seen in Fig. \ref{fig:FS_fig4}, the first few oscillations do not change 
substantially as one goes from 4/9 to 5/11 to 6/13. Therefore,
we attempt to estimate the $g(r)$ of the composite fermi sea by
fitting the first few oscillations by the function
\begin{equation}
g(r)-1=A (rk_F)^{-\alpha} \sin(\beta rk_F - \phi)\;,
\label{fit}
\end{equation}
which provides a reasonable approximation to the 
actual $g(r)$; see Fig. \ref{fig:FS_fig6} for a typical example.   
In all cases, we find that $\beta$ is close to 2 (equal to 2 within
the uncertainty in the fit), as anticipated in the above discussion. 
The presence of these $2k_Fr$ oscillations and a power-law decay of
the amplitude again provides a clear evidence
for the existence of a Fermi sea of composite fermions.
The exponent $\alpha$ is found to be smaller than 3 but 
larger than 2 (recall that $\alpha=3$ for the non-interacting Fermi 
sea); its value depends somewhat on the range over which the 
fit is attempted, so should not be taken literally. 

It is natural to ask if the pair correlation function of composite
fermions looks like that of some kind of {\em interacting} electron
liquid at zero magnetic field. Interacting fermion states are often
described by Jastrow wave functions; to make the analogy to composite
fermions as close as possible, we try the following Jastrow 
wave function for {\em interacting} electrons at zero magnetic 
field:
\begin{equation}
\Phi^{int.-FS}=\prod_{j<k}|{\bf r_j}-{\bf r}_k|^2 \Phi^{non-int.-FS} \;.
\end{equation}
The pair correlation function of this state is precisely that 
of the {\em unprojected} composite fermion wave function
at $\nu=1/2$. Motivated by this, we have  computed $g(r)$ 
for the unprojected CF state at $\nu=6/13$, also shown in 
Fig. \ref{fig:FS_fig6}. While it is qualitatively 
similar to the $g(r)$ of the projected CF wave function, or also
to the $g(r)$ of the non-interacting Fermi sea, its decay is best 
fit by $\alpha\approx 2.4$. Two conclusions may be drawn: (i)
It may indeed be possible to find some wave function of interacting 
electrons at zero magnetic field, possibly of the form
\begin{equation}
\Phi^{int.-FS}=\prod_{j<k}|{\bf r_j}-{\bf r}_k|^{x} \Phi^{non-int.-FS} \;,
\end{equation}
which would behave similar to composite fermions insofar as the 
pair correlation function is concerned. (ii) The exponent $\alpha$
is not universal; it depends on perturbations, e.g., LL mixing
(since the unprojected CF wave function simulates LL mixing to 
an extent \cite {Bonesteel}).   

\subsection{Static structure factor}

In the spherical geometry, the total angular momentum
$L$ replaces the wave vector, and spherical harmonics are 
used to define Fourier transforms. The 
static structure factor is given by (for $L\neq 0$) 
\begin{equation}
S_L=\frac{4\pi}{N}<|\sum_j Y_{L,0}(\Omega_j)|^2>
\end{equation}
where the $z$ component of $L$ is taken to be zero without 
any loss of generality. The relation $L=kR$, where $R=\sqrt{Q}l$ 
is the radius of the sphere in units of the magnetic length, 
will be used to 
present the results as a function of the wave vector $k$. 
The static structure function can be computed either directly
from the wave function, or by a Fourier transform of $g(r)$.
We have confirmed that both are in agreement; below we give
results obtained by a Monte Carlo evaluation of $S(k)$ from 
the above equation. 

\begin{figure}
\centerline{\psfig{file=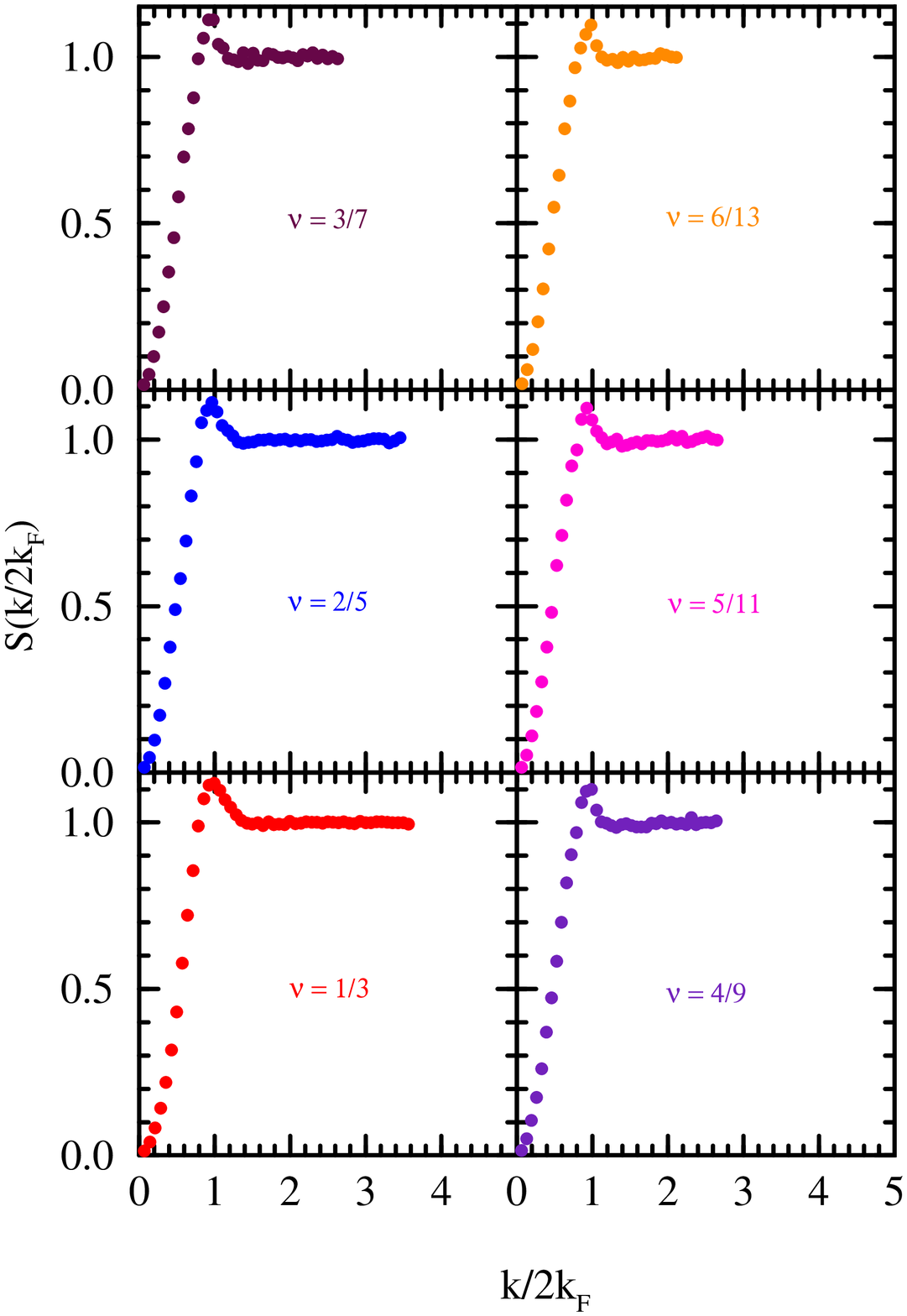,width=6in,angle=0}}
\caption{Static structure factors for the systems of Fig. \protect \ref{fig:FS_fig2}.}
\label{fig:FS_fig7}
\end{figure}

\begin{figure}
\centerline{\psfig{file=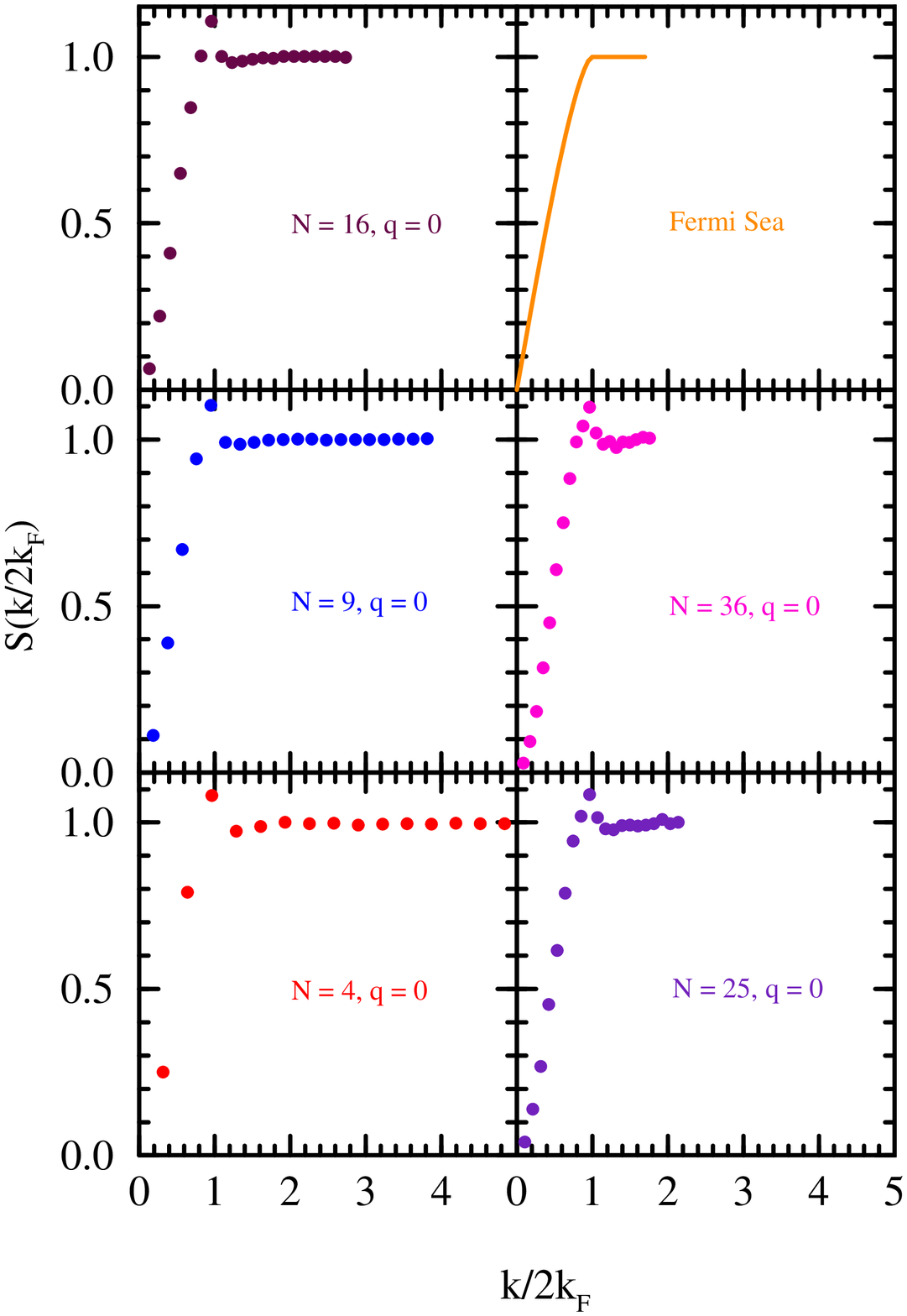,width=6in,angle=0}}
\caption{Static structure factors for the systems of Fig. \protect \ref{fig:FS_fig3}.}
\label{fig:FS_fig8}
\end{figure}

The static structure factors for the systems of Fig. \ref{fig:FS_fig2}  and
\ref{fig:FS_fig3} are plotted in Figs. \ref{fig:FS_fig7}  and \ref{fig:FS_fig8}.
Two aspects are noteworthy: $S(k)$ are all quite similar, even 
for particles as few as $N=4$, and contain  
a sharp structure at $k=2k_F$, also reported by Haldane \cite
{Haldanehalf}. It would be of interest to know if there is a 
real cusp at $k=2k_F$. This is related to the exponent $\alpha$ 
characterizing the power-law fall off of $g(r)$. A value
of $\alpha=2$ would produce a cusp in $S(k)$, as seen by noting
that the Fourier transform of $r^{-2} \sin(2k_Fr)$ is equal to
$\pi/2$ for $k<2k_F$ and arccosec$(k/2k_F)$ for $k>2k_F$.
However, no cusp will occur for $\alpha > 2.0$, suggesting
that there may not
be a real cusp in $S(k)$ at $k=2k_F$ but only a sharp peak.
The situation is complicated by the fact that only discrete
points are available for $S(k)$.

The dynamic susceptibility $\Pi(k,\omega)$ of composite fermions 
has been considered within the framework of the Chern-Simons 
description in Ref. \cite {Altshuler}, which finds a divergent
behavior as $\omega\rightarrow 0$ at $k=2k_F$, but ordinary
Fermi-liquid type behavior away from $k=2k_F$. This is 
qualitatively consistent with our results, as a divergence 
in $S(k,\omega)$ will produce an enhancement of $S(k)$ 
at wave vectors below, but close to $2k_F$, while for 
$k>2k_F$, $S(k)$ would revert back to Fermi-sea like behavior.
However, the divergence found in \cite {Altshuler}
is rather weak, and further analysis of this issue will
be useful. 

\section{Conclusion}

Using a modified variational wave function which is easily 
projected onto the lowest Landau level, 
we have computed  the pair correlation function and the static 
structure factor for composite fermions, 
and found clear evidence for Fermi-sea-like correlations in the
vicinity of the half-filled LL, making a compelling case for
the existence of a  Fermi sea of
composite fermions at $\nu=1/2$. 

The authors are grateful to P.A. Lee and  A. H. MacDonald for 
useful conversations. This work was supported in part by the 
John Simon Guggenheim Foundation (JKJ), the National
Science Foundation under grants no. DMR-9615005, DMR-9416906,
and the MRSEC Program  of the National
Science Foundation under award number DMR94-00334.

\end{document}